\documentclass[11pt,a4paper,english]{article}
\usepackage{graphicx}
\usepackage{color}
\usepackage{natbib}
\begin{document}
\title{Regularization methods for finding the relaxation time spectra of linear polydisperse polymer melts}
\author{C. Lang
}                     
%
%
%
%
\maketitle
\section*{Abstract}
The calculation of discrete or continuous relaxation time spectra from rheometric measurables of polydisperse polymers is an ill-posed problem. In this paper, a curve fitting method for solving this problem is presented and compared to selected models from the literature. It is shown that the new method is capable of correctly predicting the molecular mass distributions of linear polydisperse polymer melts as well as their relaxation time spectra.
\section{Introduction}
\label{intro}
The relaxation time spectrum (RTS) of a polymer melt is a unique representation of the underlying particle dynamics within different time regimes. Due to its molecular mass distribution (MMD), the relaxation time spectrum of a polydisperse polymer melt is more complex than the RTS of a monodisperse species. In order to account for the smeared-out representation of particle dynamics in such a spectrum, mixing rules (MRs) for the relaxation strengths and relaxation times are needed. \par
A RTS is usually calculated from rheometric measurables by conducting a rheometrical test in the quasi-linear rheological response regime over the largest possible inverse time span and applying the generalized Maxwell model to either the real or the imaginary part of the complex shear modulus. Since both parts of the complex modulus in the generalized Maxwell model are Kramers-Kronig relations \citep{Kronig}, \citep{Kramers}, their inversion is a mathematcally ill-posed problem (IPP).\par
As for every IPP, a solution can be found by using an appropriate regularization scheme \citep{Orbey}, \citep{Mead}, \citep{Roths}. A fairly simple regularization scheme is the use of a predefined response function with a reduced number of parameters as compared to the original problem. For the case of RTS, however, the definition of a response function, which was introduced as a so called MR before, is a very difficult task. This is attributed to the fact that a MR has to contain the full spectrum of particle dynamics of the polydisperse polymer melt. \par
In this paper, the above outlined scheme is used together with a curve fitting method to  calculate the RTS of an exemplary linear polydisperse polymer melt from its frequency dependent storage modulus with regards to its MMD. The method is based on a theory developed in an earlier paper \citep{Lang2}, where limitations and applicability to different polymer types was discussed in detail. The scheme is compared to existing numerical methods from the literature which use different MRs. First, the theoretical background of the mathematical and physical problem is outlined and a solution is derived. Then, the solution is compared to existing models and the MMD, which presents the only way of comparing the calculation output to measured data.

\section{Theory}
\label{sec:1}
The dynamic moduli $G'(\omega)$ and $G''(\omega)$ of a polydisperse polymer melt are functions of the angular frequency $\omega$. In order to calculate the RTS, $\{g_i,\tau_i\}$ in discrete representation and $H(\tau)$ in continuous representation, it is possible to use either of the measured moduli. Here, the storage modulus $G'(\omega)$ is used. The generalized Maxwell model leads to a Kramers-Kronig relation for the storage modulus, incorporating either the discrete RTS:
\begin{equation}
\label{KKR1}
G'(\omega)=\sum_{i=1}^Ng_i\frac{(\tau_i\omega)^2}{1+(\tau_i\omega)^2}~~,
\end{equation}
or the continuous RTS:
\begin{equation}
\label{KKR2}
G'(\omega)=\int_0^\infty d\tau H(\tau)\frac{\tau\omega^2}{1+(\tau\omega)^2}~~.
\end{equation}
The inversion of either equation~\ref{KKR1} or \ref{KKR2} would directly give the RTS. However, virtually infinite combinations of $\{g_i,\tau_i\}$ or $H(\tau)$ lead to the same outcome for the storage modulus. This means that solutions to these equations are not unique, making the inversion of both equations ill-posed according to the second Hadamard criterion \citep{Hadamard}.
In the following, five different ways to overcome ill-posedness will be presented, all of which incorporate different physical backgrounds in the predefinition of a solution to equations~\ref{KKR1} and \ref{KKR2}.\par

\subsection{Tikhonov regularization}
\label{sec:1.1}

One can calculate the RTS without using any additional physical model by inverting equations~\ref{KKR1} and \ref{KKR2} using a mathematical regularization scheme. The method by \citet{Orbey}, in the rest of the paper referred to as OD method, is shown as one example of such a scheme which uses Tikhonov regularization \citep{Tikhonov}.\par
The normal form of equation~\ref{KKR1} reads $\lim_{\delta\to0}$:
\begin{equation}
\sum_{j=1}^M\|\frac{1}{G'(\omega_j)}g_i\frac{(\tau_i\omega_j)^2}{1+(\tau_i\omega_j)^2}-1\|^2=\delta~~,
\end{equation}
where $\|\cdot\|$ stands for a vector norm. One can define a linear operator:
\begin{equation}
K_{ij}:=\frac{1}{G'(\omega_j)}g_i\frac{(\tau_i\omega_j)^2}{1+(\tau_i\omega_j)^2}~~,
\end{equation}
and use singular value decomposition $K=U\cdot W\cdot V^T$, where the superscript T stands for transposed, in order to rewrite equation~\ref{KKR1} as $\lim_{\delta\to0}$:
\begin{equation}
\label{min}
\delta=\|Kg-b\|^2=\|WV^Tg-U^Tb\|^2=\sum_{i=1}^N\left(w_iz_i-u_i^Tb\right)^2+\sum_{i=N+1}^M(u_i^Tb)^2~~,
\end{equation}
which has the minimum solution $g=Vz^\dagger$ with $z_i^\dagger=u_i^Tb/w_i$ for $i=\overline{1,N}$ and $z_i^\dagger=0$ for $i=\overline{N+1,M}$. Since this solution is arbitrary due to the above mentioned violation of the second Hadamard criterion, one can use Tikhonov regularization to find an appropriate physically meaningful solution of equation~\ref{KKR1}. This is carried out by appending a Lagrange multiplier $\lambda$ to equation~\ref{min}, such that $\lim_{\delta\to0}$:
\begin{equation}
\label{min1}
\delta=\lambda\|g\|^2+\|Kg-b\|^2~~,
\end{equation}
which possesses the solution $g=Vz_\lambda^\dagger$ with $z_\lambda^\dagger=w_iu_i^Tb/(w_i^2+\lambda)$. The Lagrange multiplier can be determined using $\lim_{\delta\to0}$:
\begin{equation}
\label{min2}
\delta=\sum_{i=1}^N\frac{w_i^2}{(w_i^2+\lambda)^2}(u_i^Tb)^2+2\|b\|^2\sum_{i=1}^N\frac{w_i^2}{w_i^2+\lambda}~~.
\end{equation}

\subsection{Heuristic regularization}
\label{sec:1.2}

Another way of regularizing the inverse problem of equation~\ref{KKR2} without using an additional physical model has been implicitly introduced in an earlier paper \citep{Lang2}. As explained above, one can use any mathematical function which consists of a small enough number of free parameters in order to regularize an IPP, as long as it represents the expected outcome within the accessible domain of the corresponding variable. 
One can fit the Cole-Cole function \citep{CC}:
\begin{equation}
\label{CC}
\ln(G'(\omega))=A+\frac{B}{1+\exp[C+D\ln(\omega)]}~~,
\end{equation}
where A, B, C and D are fitting parameters, to the experimental outcome. Since the inverse Fourier Laplace transform of equation~\ref{KKR2}, which would lead to a closed form of the RTS, is hard to obtain despite the use of this function, one uses the Stieltjes transform instead \citep{Tschoegl}, \citep{Friedrich1}. This procedure leads to the following form of the RTS:
\begin{equation}
\label{RtsColeCole}
H(\tau)=\frac{2}{\pi}\exp[f(\tau)]\sin[g(\tau)]~~,
\end{equation}
where 
\begin{equation}
f(\tau)=A+\frac{B[\exp[-C-D\ln(\tau)]+\cos(D\pi/2)]}{\exp[-C-D\ln(\tau)]+\exp[C+D\ln(\tau)]+\cos(D\pi/2)}~~,
\end{equation}
and 
\begin{equation}
g(\tau)=\frac{B\sin(D\pi/2)}{\exp[-C-D\ln(\tau)]+\exp[C+D\ln(\tau)]+\cos(D\pi/2)}~~.
\end{equation}

\subsection{Heuristic mixing rule}
\label{1.3}
The simplest physically meaningful regularization method for the inversion of equation~\ref{KKR1} stems from Schausberger \citep{Schausberger}. It is based on the BSW spectrum \citep{BSW} and will be referred to as BSW method for the rest of the paper.\par
Since the RTS of a polydisperse polymer depends on its MMD, a MR is needed to explain how the relaxation time $\tau_i$ of a polymer species with a certain mass $m_i$ is altered by the presence of another species with mass $m_j$ and relaxation time $\tau_j$ and vice versa.\par
As a first step, one creates a MMD  $\psi(m)$ with a predefined form. Here, the multimodal generalized exponential (GE) distribution 
 is used:
\begin{equation}
\label{mmd}
\psi(m)=\frac{1}{N}\sum_{i=1}^N\gamma_im^{a_i}\exp[-b_im^{c_i}]~~,
\end{equation}
where $\{a\}$, $\{b\}$, $\{c\}$ and $\{d\}$ are fitting parameters and $\gamma$ is a normalization constant.
 Then, the different masses $\{m\}$ are used to define a domain for the relaxation times $\{\tau\}$ by using the connection between the longest relaxation time of a species and its molecular mass:
\begin{equation}
\label{tauM}
\tau=\zeta m^\alpha~~,
\end{equation}
where $\zeta$ is the molecular friction coefficient and $\alpha=3.4$ is taken as constant \citep{Schausberger}. The MR for the relaxation times $\{\tau_m\}$ in the mixture is defined as:
\begin{equation}
\tau_{i,m}=\zeta m_i^\alpha\left(\sum_j\psi_jm_{j,m}/m_{i,m}\right)^{\alpha-2}~~,
\end{equation}
where $m_{i,m}=m_i\left(\sum_j\psi_jm_{j,m}/m_{i,m}\right)^{1-2\alpha}$ is called the effective molecular mass. The corresponding relaxation strengths $\{g_m\}$ of the mixture are calculated as:
\begin{equation}
g_{i,m}=\psi_i^2g_i+2\sum_{j=i+1}^N\psi_j\psi_ig_i~~,
\end{equation}
where one can use a distribution $g_i=G_0\psi_i^2$ as input set. $G_0=0.8\rho RT/m_eb_c$ is the plateau modulus, where $\rho$ is the polymer density in the melt state, $R$ is the gas constant, $T$ the absolute temperature of the melt, $m_e$ the molecular mass of entanglements and $b_c=1-\sum_i\psi_i^2$. This combination of mixing rules for the relaxation times and strengths describes the RTS of any mixture of sufficiently sharp distributed polymer species. Since for every sharply distributed species the RTS is BSW distributed \citep{BSW} one has to take into account that all relaxation times have to be weighted as:
\begin{equation}
\tau_{i,j}=\tau_{i,m}\left(\frac{\tau_i}{\tau_{i+1}}\right)^j~~,
\end{equation}
and all strengths as:
\begin{equation}
g_{i,j}=g_{i,m}\left(\frac{\tau_i}{\tau_{i+1}}\right)^{jn_e}~~,
\end{equation}
where $n_e$ describes the slope of the RTS of a monodisperse species in the rubbery regime. 
It is clear that the outcome $\{g_i,\tau_i\}$ of this procedure can be inserted in equation~\ref{KKR1} and subsequently fitted to the measurement. The input MMD is varied iteratively until the mean square deviation between the measured data and the theoretical outcome is minimized.

\subsection{Full mixing rule}
\label{1.4}
A full physical model has been presented by Carrot and Guillet \citep{Carrot} and is referred to as CG method for the rest of the paper. The deGennes-Doi-Edwards \citep{deGennes}, \citep{Doi} tube model is used to derive the relaxation strengths of a monodisperse polymer melt, starting from the Smoluchowski equation for the molecule dynamics. The Smoluchowski equation simplifies to a one dimensional diffusion equation by the use of the tube model, which possesses a solution directly proportional to the relaxation strength. This relaxation strength is, of course, alternated by the presence of surrounding chains having a certain MMD \citep{Tsenoglou}, \citep{desCloizeaux}. Depending on the masses of the surrounding chains, the polymer dynamics are entirely different. Four cases can be distinguished. If the molecular mass is lower than the critical molecular mass $m_c=2m_e$, the relaxation times depend on the molecular mass as:
 \begin{equation}
 \tau_{i,r}=\frac{6}{\pi^2}\frac{(2m_e)^{\alpha-1}}{G_0m_e}m_i^2~~.
 \end{equation} 
 The corresponding relaxation strengths are Rouse distributed \citep{Rouse}, such that:
 \begin{equation}
 g_{i,r}=\frac{\pi^2}{6}m_eG_0\psi_i/m_i~~.
 \end{equation}
If the mass equals $m_e$, one has:
\begin{equation}
\tau=\frac{\zeta N^3d^4}{\pi^2k_BTa^2}~~,
\end{equation}
where $N$ is the segment number of the polymer, $d$ is the diameter and $a$ is the step length of the primitive chain. 
The relaxation strength for such a species is:
\begin{equation}
g=\frac{\pi^2}{6}G_0\sum_i\psi_i~~.
\end{equation}
 For molecules of length higher than $m_e$, two cases can be distinguished. Either the surrounding of a chain is monodisperse, then:
 \begin{equation}
 \label{start}
 \tau_i=\tau_{i,r}\left(1+ \tau_{i,r}\left[\frac{1}{\tau_{i,m}}-\frac{1}{\tau_{i,b}}\right]\right)^{-1}~~,
 \end{equation}
 where 
 \begin{equation}
 \tau_{i,m}=3\sum_{j,k,l}\psi_i\psi_j\psi_k\frac{\tau_{j,0}\tau_{k,0}\tau_{l,0}}{\tau_{j,0}\tau_{k,0}+\tau_{k,0}\tau_{l,0}+\tau_{j,0}\tau_{l,0}}\left(\frac{m_i}{m_e}\right)^2~~,
 \end{equation}
  \begin{equation}
  \tau_{i,0}=\zeta m_em_i^{\alpha-1}~~,
  \end{equation}
  and 
  \begin{equation}
  \tau_{i,b}=\tau_{i,0}\left(\frac{m_i}{m_e}\right)^2~~.
  \end{equation}
   The corresponding relaxation strengths are:
  \begin{equation}
  \label{DoubleReptation}
  g_i=G_0\psi_i^2~~.
  \end{equation}
  Or the surrounding is polydisperse, in which case:
  \begin{equation}
\tau_i=\frac{2\tau_{i,m,j}\tau_{i,m,k}}{\tau_{i,m,j}+\tau_{i,m,k}}~~,
  \end{equation}
  where one allows for two species $j$ and $k$ with different $\tau_{i,m}$, with the corresponding strengths:
  \begin{equation}
  \label{end}
  g_i=2G_0\psi_{i,j}\psi_{i,k}~~.
  \end{equation}
  This incorporates the full modern understanding of polymer dynamics into the calculation of the RTS from a previously unknown MMD. Again, as in the BSW method, the process of RTS generation is iterated until the corresponding storage modulus fits the measurement outcome. 
  
  \subsection{New solution}
  \label{1.5}
  An attempt of simplifying the full mixing rule can be readily found in the literature \citep{Maier}. In this MR, the functional form of the relaxation strength in the tube model is nonlinearly superimposed with a distribution of molecular masses, similar to the double reptation concept, see equation~\ref{DoubleReptation} \citep{Tsenoglou}, \citep{desCloizeaux}.
  However, a difference is that the contribution of Rouse-like modes is considered by taking the difference between the actual molecular mass and the molecular mass of entanglements into account. Additionally, the mixing rule is allowed to take a fractal form. An expression introduced by  \citet{Thimm1} gives the following form for the RTS dependent MMD:
  \begin{equation}
  \label{TFMH}
  H(\tau)=\lambda^{-1}\tilde{\psi}(\tau)~~,
  \end{equation}
  where $\tilde{\psi}(\tau)=\psi(\tau(m))=\psi(m)$, 
  \begin{equation}
  \lambda=\frac{1}{\beta}\left(\frac{\alpha}{G_0}\right)^{1/\beta}\left[\int_{m_e}^\infty dmm^{-1}H(\tau)\right]^{1/\beta-1}~~,
  \end{equation}
  and $\beta$ is called the fractal exponent. This MR can now be combined with the MMD given in equation~\ref{mmd} \citep{Lang2}. Since the resulting equation for the RTS is still self contained, rewriting is necessary in order to obtain an explicit expression for the RTS of polydisperse polymers:
  \begin{equation}
  \label{analytical}
  H(\tau)=\mu G_0\sum_{i=1}^N\gamma_i\left(\frac{\tau}{\zeta}\right)^{a_i/\alpha}\exp\left[-b_i\left(\frac{\tau}{\zeta}\right)^{c_i/\alpha}\right]~~,
  \end{equation}
  where 
  \begin{equation}
  \mu=\left(\frac{1}{N}\right)^\beta\beta^{\beta+1}\frac{1}{\alpha}\left(\int_\tau^\infty d\tau'\sum_{i=1}^N\gamma_i\left(\frac{\tau'}{\zeta}\right)^{a_i/\alpha}\tau'^{-1}\exp\left[-b_i\left(\frac{\tau'}{\zeta}\right)^{c_i/\alpha}\right]\right)^{\beta-1}~~.
  \end{equation}
  The derivation of equation~\ref{analytical} is given in the appendix. All of the occurring parameters can be determined by fitting the following expression of the storage modulus to the measurement result \citep{Lang2}:
  \begin{equation}
  \label{gPrime}
  G'(\omega)=\frac{1}{N}\sum_{i=1}^N\frac{\tilde{\lambda}_i}{\nu_i}\Gamma\left(\frac{1}{\nu_i}\right)\mathcal{L}_{1/\nu_i}(b_i\zeta^{-\nu_i},0)[f'(\tau,\omega)]~~,
  \end{equation} 
  with
\begin{equation}
\label{fPrime}
f'(\tau, \omega)=\frac{\omega^2\tau}{1+(\tau\omega)^2}~~,
\end{equation}  
 where $\tilde{\lambda}=\gamma\zeta^{1/\nu}/\lambda$, $\nu=c/\alpha$, $\Gamma(s)$ is the Euler gamma function and the incomplete Laplace integral \citep{Temme}:
   \begin{equation}
   \mathcal{L}_\lambda(s,\alpha)[f(x)]=\frac{1}{\Gamma(\lambda)}\int_{\alpha}^{\infty}dxx^{\lambda-1}f(x)\exp[-sx]
   \end{equation}
   is used. The fitting function for the loss modulus reads:
   \begin{equation}
\label{gdPrime}
G''(\omega)=\frac{1}{N}\sum_{i=1}^N\frac{\tilde{\lambda}_i}{\nu_i}\Gamma(\frac{1}{\nu_i})\mathcal{L}_{1/\nu_i}(b_i\zeta^{-\nu_i},0)[f''(\tau,\omega)]~~,
\end{equation}
with 
\begin{equation}
\label{fdprime}
f''(\tau,\omega)=\frac{\omega\tau^2}{1+(\tau\omega)^2}~~.
\end{equation}
Together with the storage modulus, also the loss factor $\tan\delta=G''/G'$ of the polymer melt can be calculated.
Since the parameter space of equation~\ref{gPrime} is generally leading to ill-posedness of the problem, one confines the mode number N to the smallest integer number necessary. Here, N=3 is used in accordance with earlier findings \citep{Lang2}.
   The prefactor $\lambda$ is contained in the definition of the storage modulus, equation~\ref{gPrime}. Therefore, a fitting procedure must contain a first estimate of this parameter which requires rough knowledge of the RTS. This problem is solved here by using the heuristic regularization procedure, given in section~\ref{sec:1.2}  as a first guess and then iteratively re-evaluating the factor $\lambda$ with the new form of the RTS, given in equation~\ref{analytical}, when fitting equation~\ref{gPrime} to the measurement outcome.\par
   
  \section{Measurements and Material}
  \label{sec:2}
  \subsection{Measurements}
  \label{sec:2.1}
  Small amplitude oscillatory shear experiments were conducted on a strain controlled Anton Paar MCR 302 rheometer at the Johannes Kepler University Linz, Austria. The sample geometry was a cone-plate assembly with a cone angle of 1.5$^o$ and 25~mm diameter. The measurement temperature was 190$^o$C. The samples were pressed from granular stock material using a H\"ofer laboratory press with a 25~mm diameter punch-hole.\par
The MMD was determined by using a Viscotec high temperature gel permeation chromatography (HT-GPC) setup with a threefold column at a temperature of 140$^o$C, combined with triple detection. The setup was calibrated using a 99 ~kDa polystyrene standard. Samples were prepared, dissolving 20~mg polymer in 10~ml 1,2,4-trichloro-benzene and storage at measurement temperature for 45~min.

\subsection{Material}
\label{sec:2.2}
A commonly available batch of polydisperse polypropylene, called RA130E, by Borealis AG was used in this study. The material data sheet can be readily found online \citep{link1}.\par
Since a many-material study for the given analytical method has been presented earlier \citep{Lang2}, only this material is used here as an exemplary linear polydisperse polymer.
\section{Methods}
\label{sec:3}
In section~\ref{sec:1}, the methods of calculation are outlined to great extent, however, computation is not feasible without presetting some parameters. Many parameters used in the common methods are not obtainable through experimental investigation of industrial polymer grades.\par
In table~\ref{tab:1}, the important a priori parameters to all methods using mixing rules are given. It has to be emphasized that most values do not stem from experimental evidence, but have been obtained by preliminary numerical studies \citep{Thimm1, Lang1}.
\begin{table}
	\caption{Choice of a priori parameters.}
	\label{tab:1}       
	\begin{tabular}{lll}
		\hline\noalign{\smallskip}
		name & value & source  \\
		\noalign{\smallskip}\hline\noalign{\smallskip}
		$\zeta$ & 3x10$^{-21}$~s/g & \citet{Eder} \\
		$m_e$ & 4390~g/mol & \citet{Lang1} \\
		$n_e$ & 0.23 & \citet{Schausberger} \\
		$\beta$ & 3.84 & \citet{Lang1} \\
		\noalign{\smallskip}\hline
	\end{tabular}
\end{table}

\section{Results}
\label{sec:4}
\begin{figure}[h]
	\begin{center}
		\includegraphics[scale=0.3]{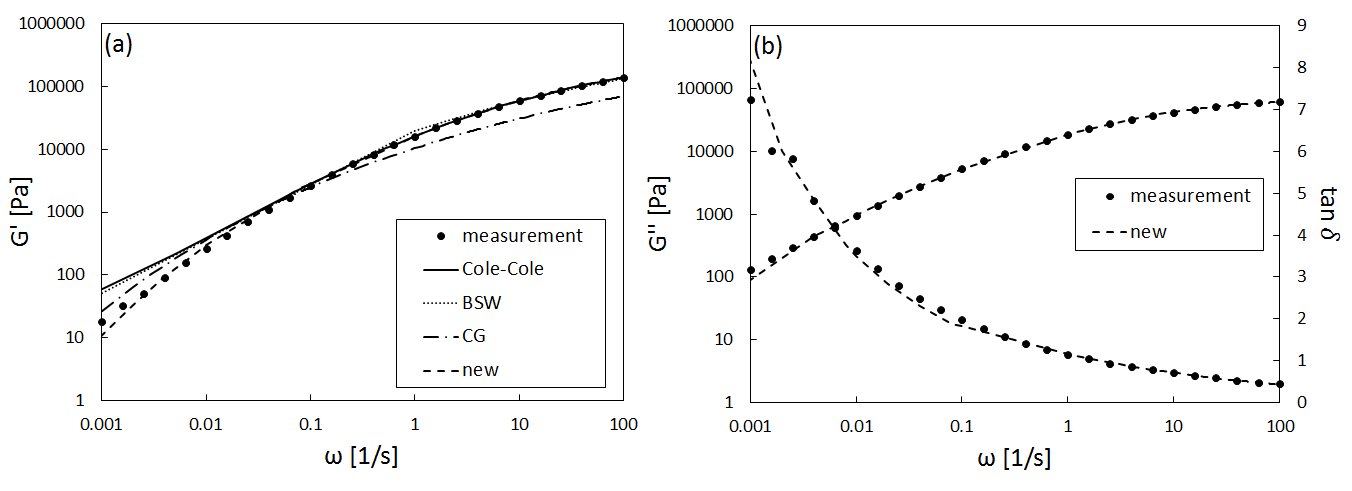}
		\caption{(a) Storage modulus as a function of frequency, compared to different theoretical fits. (b) Loss modulus and loss factor of the new method compared to the experimental outcome.}
		\label{fig:1}
	\end{center}
\end{figure}

Since the measurement of the storage modulus and its approximation by the different theoretical methods lie at the heart of the given problem, they are shown in figure~\ref{fig:1}(a). In addition, fitting results for the loss modulus and loss factor obtained from the new method are shown in figure~\ref{fig:1}(b).

It is clear that methods capable of correctly fitting the measurement, in general, have a higher chance to give correct MMDs and RTS. However, no direct correlation between a fitting result and the correctness of the outcome exists, as is discussed in section~\ref{sec:5}.\par
Since the regularization procedure of the OD method uses the measurement as input, it is not included in figure~\ref{fig:1}(a). \par
\begin{figure}[h]
	\begin{center}
		\includegraphics[scale=0.3]{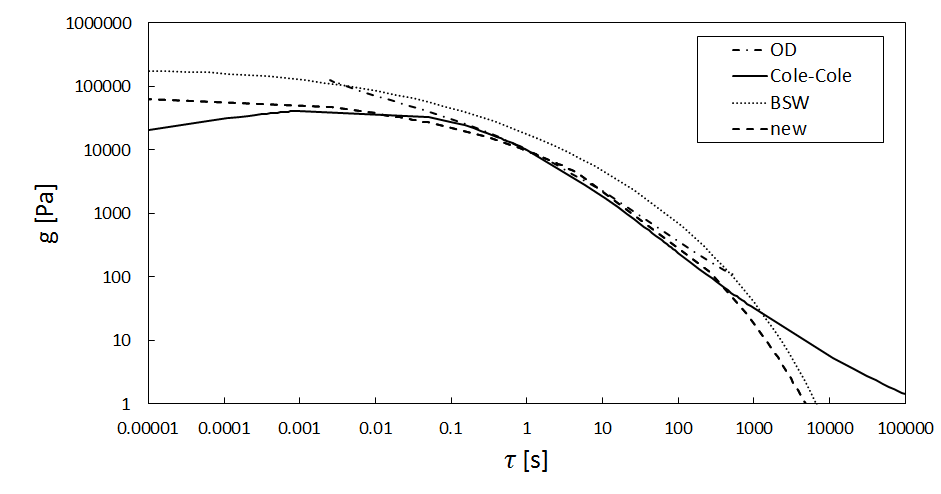}
		\caption{Relaxation time spectra from regularization, compared to the new method.}
		\label{fig:2}
	\end{center}
\end{figure}

Regularization alone, by means of the OD method or the heuristic regularization, results in RTS for the given material. A comparison of results, including also the BSW and the new solution, is shown in figure~\ref{fig:2}.

It can be seen that the curve of the OD model is not spanning the entire interesting range of relaxation times. This feature is discussed in section~\ref{sec:5}.\par
\begin{figure}[h]
	\begin{center}
		\includegraphics[scale=0.3]{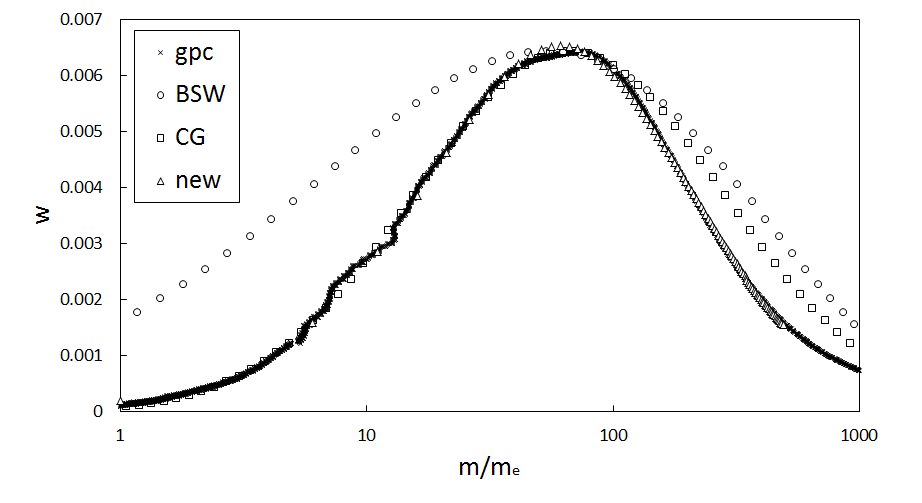}
		\caption{Molecular mass distribution obtained from different mixing rules compared to HT-GPC data.}
		\label{fig:3}
	\end{center}
\end{figure}
\begin{figure}[h]
	\begin{center}
		\includegraphics[scale=0.3]{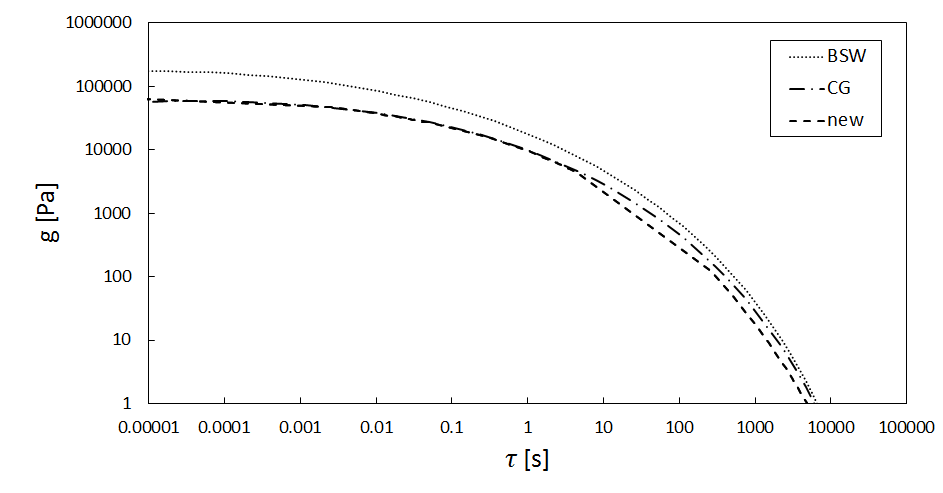}
		\caption{Relaxation time spectra from methods based on mixing rules.}
		\label{fig:4}
	\end{center}
\end{figure}

During the approximation of the storage modulus, the theoretical methods based on mixing rules update both, the MMD and the RTS, until the error $\delta$ between the measured data points and the calculated storage modulus is minimized. The resulting MMDs are shown in figure~\ref{fig:3} and the corresponding RTS are presented in figure~\ref{fig:4}. 

\section{Discussion}
\label{sec:5}
From figure~\ref{fig:1}(a) it is evident that the new curve fitting procedure gives the best overall fit to the measured storage modulus of the industrial polypropylene grade. Figure~\ref{fig:1}(b) shows that also the loss modulus and loss factor are approximated well by the new method. The fits obtained from the BSW and Cole-Cole methods are reasonably good in the high frequency regime, but diverge from the measurement in the low frequency regime. Surprisingly, the worst fit to the data is obtained by using the CG procedure. However, since the CG curve for the storage modulus lies above the measurement at low frequencies and below at high frequencies, the shown result nonetheless represents a minimal error $\delta$. The errors of all shown fits are given in table~\ref{tab:2}.
 
\begin{table}
	\caption{Errors of the storage modulus fit for different methods.}
	\label{tab:2}       
	\begin{tabular}{ll}
		\hline\noalign{\smallskip}
		method & error [\%]  \\
		\noalign{\smallskip}\hline\noalign{\smallskip}
		BSW & 0.22 \\
		CG & 3.57 \\
		Cole-Cole & 0.25 \\
		analytical & 0.1 \\
		\noalign{\smallskip}\hline
	\end{tabular}
\end{table}

It needs to be emphasized that a good fit to the storage modulus alone does not necessarily provide a good representation of the MMD or the RTS. If the fit of the storage modulus represents a minimal error, the corresponding solution is always the best obtainable solution for a given method. Therefore, a physical comparison of methods based on the fits to the storage modulus measurement is not meaningful. The fits are shown here for the sake of completeness only and they can be used during application of a method to test whether a minimal error was obtained.\par
A first comparison of method outcomes can be seen in figure~\ref{fig:2}, where the RTS predicted by the OD and Cole-Cole method are compared to the new solution as well as the BSW method. One can see that the OD method describes the RTS well at intermediate relaxation times, but diverges from the novel solution at higher and lower relaxation times. The Tikhonov regularized spectrum lies between the BSW and the new spectrum. However, large parts of the spectrum are not covered by the OD method. The reason is that the OD method loses its constraints at low as well as high relaxation times, corresponding to $\omega^{-1}$ values beyond the measurement range for $G'(\omega)$. Since the inverse of the measured range of $\omega$ is physically not directly correspondent to the range of relaxation times for which a model can be used, one can identify this as a disadvantage of the OD method as compared to other methods. \par
In contrast to the OD method, the Cole-Cole regularization method gives a full prediction of the RTS. The outcome, however, diverges from the new as well as the BSW result at low and high relaxation times, underestimating the relaxation strengths at low relaxation times and overestimating the strengths at high relaxation times. The high relaxation time mismatch can be explained by a misfit of the measured storage modulus curve in the low frequency regime, as is seen from figure~\ref{fig:1}(a). In the low relaxation time regime, the difference in RTS between the Cole-Cole regularization and the new solution most likely results from missing constraints for the Cole-Cole fitting procedure. \par
Since the novel method relies on a robust first estimate of the RTS in the full interesting relaxation time regime, the heuristic regularization method is chosen for this purpose.\par
Figure~\ref{fig:3} shows a comparison of the MMDs obtained from different methods to the HT-GPC measurement. It is evident that the result of the BSW method is profoundly different from the measured MMD. The MMD obtained via the CG method, in contrast, gives a very good estimate for low molecular masses and overestimates the higher masses. This is a first interesting observation in the sense that it hints at the necessity of a full physical model in order to describe the relation between MMD and storage modulus correctly.\par
The discrepancies between the BSW and the CG method entirely result from the fact that Rouse-like modes are completely neglected in the BSW method. As low mass compounds contribute to both, the dilation of tubes for longer molecules and the importance of fast relaxation processes, they cannot be regarded as if they were moving via double reptation.\par
In the high molecular mass regime, the CG method also overestimates the occurrence of masses in comparison to the measurement. One can attribute this also to the fact that Rouse modes are more effective in tube dilation than taken into account by equations~\ref{start} to \ref{end}. In the CG model, Rouse modes contribute to those parts of the MRs insofar as the difference between relevant molecular mass and entanglement mass is considered. However, no Rouse modes are directly used to dilate the underlying double reptation.\par 
Following this train of thought, it might seem surprising that the new result describes the MMD obtained from HT-GPC well. The fact that the fractal reptation exponent holds a value higher than 2 thereby points in the direction that double reptation is not enough to describe the effect of a polydisperse surrounding on the motion of a single chain. This, however, is a slightly misleading hint, since the magnitude of the reptation exponent $\beta$ depends also on the importance of Rouse modes, as has been shown by \citet{Thimm2}. The difference between the relevant molecular mass and the molecular mass of entanglement, using the higher reptation exponent, is accounting for the importance of Rouse-like modes in the high molecular mass regime. This holds for certain materials, as has been discussed in an earlier work \citep{Lang2}. A detailed analysis of the magnitude of the reptation coefficient has been conducted by \citet{Friedrich2}. Their work also adresses an important issue of the new model, namely, the reproduction of a nearly monodisperse MMD. Due to the assumed single exponential time decay, as is usual for the generalized Maxwell model, the new result shown here does not apply to nearly monodisperse polymer melts. Particularly, the given series of exponential functions representing the MMD cannot be reduced to a delta distribution \citep{Friedrich2}.
\par
In the peak mass region of the MMD, one can observe that the new result does not cover all details of the measurement. Since the overall structure of the MMD resulting from the novel method has been discussed in detail elsewhere \citep{Lang2}, this detail is ignored here, also regarding the fact that an extended review of the new method in terms of a many-material study has been given earlier. It has been shown there that polymer melts with a multi-modal MMD can be represented by the new curve fitting results as well.\par
RTS from different methods are compared in figure~\ref{fig:4}. In comparison to figure~\ref{fig:3}, a direct correlation between MMD and RTS can be detected. This could be an indication for correctness of the analytical prediction by Thimm et al. given in equation~\ref{TFMH}. They claim that the form of the MMD is equivalent to the form of the RTS, differing only by a scaling parameter and a change in variables. This result is fully recovered by the new method.\par
The presented method provides a new way of obtaining relaxation time spectra from rheometric data of industrial grade polymers with a higher precision than other models.\par
In the novel method, unfortunately, the physical meaning of the individual parameters, resulting from a completely adaptable functional form of the MMD, as given in equation~\ref{mmd}, is lost. Despite the given differences in the mixing rule, no physical meaning of the parameters contained in the RTS of the new method can be extracted. Nonetheless, the given method is one of a few curve fitting forms available for polydisperse polymer melts \citep{Friedrich2}, all of which contain a certain number of fitting parameters. These fitting parameters are either introduced ad hoc, as in the case of the BSW RTS \citep{BSW}, or for the purpose of regularization.

\section{Conclusions}
\label{sec:6}
A new form of the MMD dependent RTS of linear polydisperse polymer melts is presented.  Despite the fact that this form relies on a simplified MR for the underlying polymer dynamics, the new method is capable of correctly predicting the MMD as well as the RTS of industrial polymer melts with a higher accuracy than other methods.

%

%
%
%

\section*{Appendix}
\label{appendix}
Starting from the self-contained form of the RTS:
\begin{equation}
H(\tau)=\frac{1}{N}\beta\left(\frac{G_0}{\alpha}\right)^{1/\beta}\left(\int_{\tau_e}^{\infty}d\tau\tau^{-1}H(\tau)\right)^{1-1/\beta}\sum_{i=1}^N\gamma_i\left(\frac{\tau}{\zeta}\right)^{a_i/\alpha}\exp\left[-b_i\left(\frac{\tau}{\zeta}\right)^{c_i/\alpha}\right]~~,
\end{equation}
one can multiply both sides with $\tau^{-1}$ and integrate over $\tau$ from $\tau_e$ to $\infty$, giving:
\begin{equation}
\label{integralH}
\int_{\tau_e}^{\infty}d\tau\tau^{-1}H(\tau)=\left(\frac{\beta}{N}\right)^\beta\frac{G_0}{\alpha}\left(\int_{\tau_e}^{\infty}d\tau\sum_{i=1}^N\gamma_i\zeta^{-a_i/\alpha}\tau^{a_i/\alpha-1}\exp\left[-b_i\left(\frac{\tau}{\zeta}\right)^{c_i/\alpha}\right]\right)^\beta~~.
\end{equation}
By recognizing that:
$$\left(\int_{\tau_e}^{\infty}d\tau\sum_{i=1}^N\gamma_i\zeta^{-a_i/\alpha}\tau^{a_i/\alpha-1}\exp\left[-b_i\left(\frac{\tau}{\zeta}\right)^{c_i/\alpha}\right]\right)^\beta$$
$$=\beta\int_{\tau_e}^{\infty}d\tau\sum_{i=1}^{N}\gamma_i\zeta^{-a_i/\alpha}\tau^{a_i/\alpha-1}\exp\left[-b_i\left(\frac{\tau}{\zeta}\right)^{c_i/\alpha}\right]$$
\begin{equation}
\times\left(\int_{\tau}^{\infty}d\tau'\sum_{i=1}^{N}\gamma_i\zeta^{-a_i/\alpha}\tau'^{a_i/\alpha-1}\exp\left[-b_i\left(\frac{\tau'}{\zeta}\right)^{c_i/\alpha}\right]\right)^{\beta-1}~~,
\end{equation}
one can rewrite equation~\ref{integralH} as:
$$\int_{\tau_e}^{\infty}d\tau\tau^{-1}H(\tau)=$$
$$\left(\frac{1}{N}\right)^\beta\beta^{\beta+1}\frac{G_0}{\alpha}\int_{\tau_e}^{\infty}d\tau\sum_{i=1}^{N}\gamma_i\zeta^{-a_i/\alpha}\tau^{a_i/\alpha-1}\exp\left[-b_i\left(\frac{\tau}{\zeta}\right)^{c_i/\alpha}\right]$$
\begin{equation}
\times\left(\int_{\tau}^{\infty}d\tau'\sum_{i=1}^{N}\gamma_i\zeta^{-a_i/\alpha}\tau'^{a_i/\alpha-1}\exp\left[-b_i\left(\frac{\tau'}{\zeta}\right)^{c_i/\alpha}\right]\right)^{\beta-1}~~.
\end{equation}
Assuming that the integral vanishes if the integrand goes to zero, equation~\ref{analytical} is obtained.

\section*{Acknowledgements}
The author wishes to acknowledge I. Teasdale from the Institute of Polymer Chemistry at the Johannes Kepler University Linz, Austria for the help in conducting and evaluating HT-GPC measurements and M. Gall for the fruitful discussions. This research is funded by the European Union within the Horizon 2020 project under the DiStruc Marie Sk\l{}odowska Curie innovative training network; grant agreement no. 641839.

\bibliographystyle{apalike}
\bibliography{RTSArx}

\end{document}